\documentclass[a4paper]{jpconf}
\usepackage{graphicx}

\usepackage[dvips]{color}
\definecolor{Black}{named}{Black}
\definecolor{Blue}{named}{Blue}
\definecolor{Red}{named}{Red}

\newcommand{\I}{{\rm i}}

\newcommand{\D}{{\rm d}}

\begin{document}

\title{\hbox to0pt{\vbox to 0pt{\vskip0pt plus10cm minus20cm
\vbox{\rm\normalsize Contribution to TAUP 2007,
 11--15 Sept.\ 2007, Sendai, Japan
 \vskip3cm}}\hfilneg}Multi-angle effects in
collective supernova neutrino oscillations}

\author{A Esteban-Pretel$^1$, S Pastor$^1$, R Tom\`as$^1$,
G G Raffelt$^2$, G Sigl$^{3}$}

\address{$^1$ AHEP Group, Institut de F\'\i sica
 Corpuscular, CSIC/Universitat de Val\`encia,\\
 Edifici Instituts d'Investigaci\'o, Apt.\ 22085,
 46071 Val\`encia, Spain}
\address{$^2$ Max-Planck-Institut f\"ur Physik
 (Werner-Heisenberg-Institut), F\"ohringer Ring 6,\\ 80805 M\"unchen,
 Germany}
\address{$^3$ II.~Institut f\"ur Theoretische Physik,
Universit\"at Hamburg, Luruper Chaussee 149,\\ 22761 Hamburg,
Germany}

\ead{andreu.esteban@ific.uv.es, sergio.pastor@ific.uv.es,
ricard@mppmu.mpg.de, raffelt@mppmu.mpg.de, sigl@mail.desy.de}

\begin{abstract}
We study two-flavor collective neutrino oscillations in the
dense-neutrino region above the neutrino sphere in a supernova (SN).
The angular dependence of the neutrino-neutrino interaction
potential causes ``multi-angle'' effects that can lead either to
complete kinematical decoherence in flavor space or only to small
differences between different trajectories. This nonlinear system
switches abruptly between ``self-maintained coherence'' and
``self-induced decoherence'' among the angular modes, depending on
the strength of the deleptonization flux. For a realistic SN the
quasi single-angle behavior is probably typical, simplifying the
numerical treatment and probably allowing for the survival of
observational features of flavor oscillations.
\end{abstract}

The neutrinos streaming off a collapsed supernova (SN) core are so
dense near the neutrino sphere that they produce a significant
refractive effect for each other, leading to collective oscillation
effects. The practical importance of these nonlinear phenomena was
only recently recognized and studied in a series of
papers~\cite{Sawyer:2004ai, Sawyer:2005jk, Duan:2005cp, Duan:2006an,
Hannestad:2006nj, Duan:2007mv, Raffelt:2007yz, EstebanPretel:2007ec,
Raffelt:2007cb, Raffelt:2007xt, Duan:2007fw, Duan:2007bt,
Fogli:2007bk, Duan:2007sh, EstebanPretel:2007yq}. An introduction was
given at this conference by Eligio~Lisi~\cite{Lisi2007}.

The SN neutrino fluxes are thought to obey the hierarchy
$F_{\nu_e}>F_{\bar\nu_e}>F_{\nu_x}$ where $\nu_x$ stands for any of
$\nu_\mu$, $\nu_\tau$, $\bar\nu_\mu$, and $\bar\nu_\tau$. In other
words, there is an excess of $\nu_e\bar\nu_e$ pairs relative to
$\nu_x\bar\nu_x$. In an inverted hierarchy situation, this pair
excess converts collectively into $\nu_x\bar\nu_x$ pairs, a process
that does not violate flavor-lepton number and thus does not require
mixing: it could also proceed as an ordinary pair annihilation
process. Neutrino refraction causes this pair process to proceed
collectively and very fast, almost independently of the mixing angle,
i.e., we have to do with a collective ``speed-up
effect''~\cite{Sawyer:2004ai, Sawyer:2005jk}. The unpaired $\nu_e$
excess flux from deleptonization is
conserved~\cite{Hannestad:2006nj}. In the adiabatic limit, it is the
low-energy part of the $\nu_e$ spectrum that survives, leading to a
step-like feature in the $\nu_e$ spectrum (a ``spectral
split''~\cite{Raffelt:2007cb, Raffelt:2007xt} caused by a ``step-wise
spectral swapping''~\cite{Duan:2006an, Duan:2007bt}). All of these
effects happen in the dense-neutrino region that typically extends
from the neutrino sphere out to a few hundred kilometers. If the
ordinary MSW resonances occur at larger radii, the collective
phenomena and subsequent MSW transformations are independent, the
former producing the initial condition for the latter. For an
inverted hierarchy the only effect of matter in the
collective-transformation region is a decrease of the effective mixing
angle~\cite{Hannestad:2006nj, EstebanPretel:2007ec}. Conversely, if
the matter density profile is very shallow, the ordinary MSW effect
can prepare the initial condition for the collective
effects~\cite{Duan:2007sh}.

Mixed neutrinos are described by matrices of density $\rho_{\bf p}$
and $\bar\rho_{\bf p}$ for each (anti)neutrino mode. The diagonal
entries are the usual occupation numbers whereas the off-diagonal
terms encode phase information. The equations of motion are
$\I\partial_t\varrho_{\bf p}=[{\sf H}_{\bf p},\varrho_{\bf p}]$,
where the Hamiltonian is~\cite{Sigl:1992fn}
\begin{equation}\label{eq:hamiltonian}
 {\sf H}_{\bf p}=\Omega_{\bf p}
 +{\sf V}+\sqrt{2}\,G_{\rm F}\!
 \int\!\frac{\D^3{\bf q}}{(2\pi)^3}
 \left(\varrho_{\bf q}-\bar\varrho_{\bf q}\right)
 (1-{\bf v}_{\bf q}\cdot{\bf v}_{\bf p}),
\end{equation}
${\bf v}_{\bf p}$ being the velocity. In the mass basis, the matrix
of vacuum oscillation frequencies is $\Omega_{\bf p}={\rm
diag}(m_1^2,m_2^2,m_3^2)/2E$. The matter effect is represented, in
the weak interaction basis, by ${\sf V}=\sqrt{2}\,G_{\rm F}n_B\,{\rm
diag}(Y_e,0,Y_\tau^{\rm eff})$. For antineutrinos the only difference
is $\Omega_{\bf p}\to-\Omega_{\bf p}$. The effective tau-lepton
density $Y_\tau^{\rm eff}\approx 10^{-5}$ arises from radiative
corrections~\cite{Botella:1986wy} and can be important in a genuine
three-flavor treatment of ordinary~\cite{Akhmedov:2002zj} or
collective~\cite{EstebanPretel:2007yq} SN neutrino oscillations. For
the latter, the influence of $Y_\tau^{\rm eff}$ can be rather
sensitive to deviations from maximal 23-mixing.

The angular factor $(1-{\bf v}_{\bf q}\cdot{\bf v}_{\bf p})$ in
Eq.~(\ref{eq:hamiltonian}) derives from the current-current nature of
the weak-interaction Hamiltonian. In an isotropic ensemble it
averages to unity, whereas the neutrinos streaming off a SN core are
strongly non-isotropic so that different angular modes experience a
different strength of the neutrino-neutrino interaction potential.
One may expect that this effect leads to kinematical decoherence
among angular modes and thus to flavor
equilibrium~\cite{Sawyer:2004ai, Sawyer:2005jk}. In a symmetric
ensemble with equal densities of neutrinos and antineutrinos, this is
indeed the case. Such a system is highly unstable in that an
infinitesimal deviation from exact isotropy is enough to trigger an
exponential run-away towards flavor
equilibrium~\cite{Raffelt:2007yz}. On the other hand, a numerical
simulation of the flavor evolution of SN neutrinos revealed that
multi-angle effects were small: All angular modes evolved nearly
collectively, very similar to an isotropic
ensemble~\cite{Duan:2006an}. Likewise, in a multi-energy system every
energy mode feels a different Hamiltonian, yet the evolution is
collective, i.e., multi-energy effects do not lead to kinematical
decoherence.

We have performed a numerical exploration of multi-angle effects in a
spherically symmetric system where the neutrinos are emitted from a
``neutrino sphere''~\cite{EstebanPretel:2007ec}. We have identified
the asymmetry between the $\nu_e$ and $\bar\nu_e$ flux as the crucial
parameter. Realistic deleptonization fluxes in SNe seem sufficient to
suppress multi-angle decoherence. Therefore, in practice multi-angle
effects seem to be a subdominant feature of collective SN
transformations.

A better understanding of collective oscillations can be developed in
the two-flavor case in terms of the usual flavor polarization vectors
${\bf P}_{\bf p}$ to express the matrices $\rho_{\bf p}$. The
equations of motion are $\partial_t{\bf P}_{\bf p}={\bf H}_{\bf
p}\times{\bf P}_{\bf p}$. Ignoring the ordinary matter effect, the
``Hamiltonian'' is
\begin{equation}\label{eq:hamiltonian2}
 {\bf H}_{\bf p}=\omega {\bf B}
 +\sqrt{2}\,G_{\rm F}\!
 \int\!\frac{\D^3{\bf q}}{(2\pi)^3}
 \left({\bf P}_{\bf q}-\bar{\bf P}_{\bf q}\right)
 (1-{\bf v}_{\bf q}\cdot{\bf v}_{\bf p}),
\end{equation}
where ${\bf B}$ is a unit vector in flavor space in the ``mass
direction'' and $\omega=\Delta m^2/2E$ the vacuum oscillation
frequency. In an isotropic situation where the velocity terms average
to zero, the second term is of the form $\mu\,{\bf D}$ where the
vector ${\bf D}$ is the difference between the total neutrino and
antineutrino polarization vectors and $\mu=\sqrt2 G_{\rm F}\,n_{\nu}$
the neutrino-neutrino interaction strength. The individual ${\bf
H}_{\bf p}=\omega {\bf B}+\mu{\bf D}$ all lie in a single plane. In
the adiabatic limit the polarization vectors follow the Hamiltonians
so that they also lie in this co-rotating plane. This observation
explains the collective nature of the multi-energy evolution (all
polarization vectors stay in the same plane that rotates around ${\bf
B}$ with a certain frequency $\omega_{\rm c}$) and explains the
spectral splits in that the final ${\bf H}_{\bf p}$ in the
co-rotating plane are either aligned or anti-aligned with ${\bf
B}$~\cite{Raffelt:2007cb, Raffelt:2007xt}.  This picture also
illustrates a crucial difference to the symmetric system where
initially ${\bf D}=0$ and the overall evolution is purely
pendular~\cite{Hannestad:2006nj}---there is no co-rotating plane.

In a spherically symmetric situation every mode is characterized by
its vacuum oscillation frequency $\omega$ and its angle $\theta$
relative to the radial direction, providing $v=\cos\theta$ as the
radial velocity. Now the individual Hamiltonians are ${\bf H}_{\bf
p}=\omega {\bf B}+\mu({\bf D}-v_{\bf p}{\bf F})$ where ${\bf F}$ is
the flux term of the neutrino ensemble, i.e., the same as ${\bf D}$,
but every mode weighted with $v_{\bf p}$. If all polarization vectors
lie in a single plane, then also all ${\bf H}_{\bf p}$ are in that
plane so that an evolution in a co-rotating plane is
self-consistently possible. Our numerical simulations show that those
cases with little kinematical decoherence correspond to the
polarization vectors essentially staying in a co-rotating plane with
some zenith-angle spread. On the other hand, the vectors ${\bf B}$,
${\bf D}$ and ${\bf F}$ do not have to stay in a single plane and in
fact, this appears to be an unstable arrangement. Kinematical
multi-angle decoherence corresponds to strong deviations from this
coplanar situation. In a spherically symmetric situation, the
neutrino-neutrino interaction strength decreases with $r^{-4}$. In a
toy model with an artificially slow decrease of $\mu(r)$ one can make
the evolution arbitrarily slow and adiabatic. In this case
multi-angle decoherence appears to be unavoidable.

It appears that in a realistic SN the collective evolution is slow
enough to be essentially adiabatic with regard to the development of
a spectral split, but fast enough that the co-planar arrangement of
the polarization vectors survives. This picture may provide the key
to an analytic understanding of the conditions for multi-angle
kinematical decoherence.

The assumption of spherical symmetry of the overall system severely
restricts possible solutions. The evolution is a one-dimensional
problem along the radial direction. It remains to be studied if this
symmetry has an important impact on kinematical decoherence, i.e., if
a system with fewer symmetries would decohere more easily.

\ack

This work was partly supported by the Deutsche
Forschungsgemeinschaft (TR-27 ``Neutrinos and Beyond''), by the
Cluster of Excellence ``Origin and Structure of the Universe''
(Garching and Munich), by the European Union (contracts No.\
RII3-CT-2004-506222 and MRTN-CT-2004-503369), and by the Spanish
grants FPA2005-01269 (MEC) and ACOMP07-270 (Generalitat Valenciana).
AE~was supported by an FPU grant from the Spanish Government. SP~and
RT were supported by MEC contracts ({\em Ram\'{o}n
  y Cajal} and {\em Juan de la Cierva}, respectively).

\section*{References}

\end{document}